\documentclass[aps,prl,twocolumn,showpacs,floatfix]{revtex4}

\usepackage{graphicx}
\usepackage{color}
\usepackage{bm}

\begin{document}

\bibliographystyle{apsrev}

\title{Rayleigh-Taylor turbulence is nothing like Kolmogorov's in the
 self similar regime}
\author{\surname{Olivier} Poujade}

\affiliation{Commissariat \`a l'Energie Atomique\\
BP12, Bruy\`eres-le-Ch\^atel, 91168 France}
\date{\today}

\begin{abstract}
An increasing number of numerical simulations and
experiments describing the turbulent spectrum of Rayleigh-Taylor (RT)
mixing layers came to light over the past few years. Results
reported in recent studies allow to rule out a turbulence {\it
\`a la Kolmogorov} as a mechanism acting on a self similar RT turbulent
mixing layer. A different mechanism is presented, which complies with
both numerical and experimental results and relates RT flow to other
buoyant flows.     
\end{abstract}

 \maketitle

A Rayleigh-Taylor (RT) instability \cite{rayl,tayl} occurs whenever a light fluid, $\rho_1$,
pushes a heavy fluid, $\rho_2$, or similarly, when a heavy fluid on
top of a lighter fluid is subject to a gravitational acceleration. The
understanding of such instability in the developed turbulent regime is
of primary interest to many fields of physics and technology since it
is an important cause of mixing between two fluids of different
density. In astrophysics for instance, it is responsible for the
outward acceleration of a thermonuclear flame in type Ia supernovae
\cite{astro1}, but it also plays an important role in shaping the
interstellar medium so that new stars can be born \cite{astro2}. The
technology of confinement fusion also relies on a good understanding
of RT mixing \cite{icf2} and ways to reduce it \cite{icf1}. The RT
flow of two incompressible fluids in the low Atwood limit, ${\cal
  A}=(\rho_2-\rho_1)/(\rho_2+\rho_1)\ll 1$ (Boussinesq approximation),
is governed by a concentration equation (\ref{eqc}), the Navier Stokes
equation supplemented with a buoyant source term (\ref{equ}) and the
incompressibility constraint (\ref{eqdiv})  
\begin{eqnarray}
\partial_t\,c+\left(\bm{u}\bm{\nabla}\right)c\,&=&\kappa\,\Delta c~,\label{eqc}\\
\partial_t\,\bm{u}+\left(\bm{u}\bm{\nabla}\right)\bm{u}\,&=&-\bm{\nabla} P+2\,{\cal A}\bm{g}\,c+\nu\,\Delta\bm{u}~,\label{equ}\\
\bm{\nabla}\bm{u} &=&0~,\label{eqdiv}
\end{eqnarray} where $\bm{g}$ is a stationary and uniform
gravitational acceleration vector field (i.e. planar symmetry is
  assumed). The coefficient $\kappa$ is the molecular diffusion
  coefficient and $\nu$ is the kinematic viscosity of the mixture. They
  are both supposed constant. Without loss of
  generality, $\bm{g}$ is parallel to the $z$-axis. This
  is why, for any generic physical value $\Phi$, the average (so
  defined numerically and experimentally) will be
  $\langle\Phi\rangle(z)=\frac{1}{S}\int_S dx\,dy\,\Phi$ throughout this
  paper. The fluctuating part will be denoted with a prime and defined
  as $\Phi^{\prime}=\Phi-\langle\Phi\rangle$.

With the increasing capacity of super computers, many simulations of
RT flows in the developed turbulent regime have been performed, which
describe the velocity spectrum $E(k)$ \cite{csz,cook2,young},
defined in such a way that $\langle{\bm{u}^{\prime}}^2\rangle=\int
dk\,E(k)$, or the concentration spectrum $E_c(k)$ \cite{rama1, dim1,
  cook1, dalz1}, defined as $\langle{c^{\prime}}^2\rangle=\int
dk\,E_c(k)$, or both \cite{cab1, cook3}. In the same time, although
fewer in number, experimental investigations of $E(k)$ \cite{andr3}
and $E_c(k)$ \cite{andr1, andr2, dalz1} have been carried out. A quick
inspection of these results shows that no consensus arises concerning
the shape of these spectrum. From a theoretical point of view, the
situation is not more satisfactory. In \cite{chert1} it is claimed
that the Kolmogorov-Obukhov scheme, $E(k)\sim k^{-5/3}$, holds in 3D
RT mixing given that the effect of buoyancy on turbulence, although
fundamental at the largest scale, becomes irrelevant at smaller
scales. In \cite{zhou} the particular RT time scale $1/\sqrt{k g A}$
at wave number $k$ has been postulated to vary like the turbulent
spectral transfer time scale $\epsilon\,k^{-4}\,E^{-2}(k)$ yielding an
RT turbulent spectrum $E(k)\sim k^{-7/4}$. Another troublesome point
is that there are no convincing objective criteria to assert whether or
not an RT flow has reached the gravitational self similar regime but
to plot the mixing zone width $L(t)$
versus ${\cal A}\,g\,t^2$ and see if there is a straight line
somewhere without any information on the expected slope. Subsequently, suggesting the possibility of any behavior of
the velocity spectrum in the self similar regime and comparing it to
experiments and simulations can be dubious if it is not known for
certain that these experiment or simulations have reached this
regime. So, the first aim of this paper will be to describe
such an objective criterion. The ultimate goal will be to prove that
the Kolmogorov mechanism does not explain the observed numerical 
and experimental results. A theory, based on a spectral equation, will be 
presented which shows that a balance mechanism between buoyancy and spectral
energy transfer can settle at low wave numbers in the self similar regime. 

The RT developed turbulent regime, as complex as
 it may look, is thought to evolve self similarly
 \cite{rist}. The size of the
 largest significant turbulent structure, therefore, grows like the size of
 the turbulent mixing zone which evolves as $L(t)=\alpha\,{\cal A}\,g\,t^2$
 \cite{dim1} where $\alpha$ is the mixing-zone-width growth rate parameter
 whose value is around $\sim 0.06$ experimentally and between $0.03$
 and $0.07$ numerically. It can be assumed that at low wave numbers in the self similar regime the
 velocity spectrum $E(k,t)=0$ for $k<k_l(t)$, where $k_l(t)$ is the wave number corresponding to 
the maximum of the velocity spectrum (this point will be called $\lambda$ due to the shape of the
idealized spectrum at this location), and $E(k,t)\approx
 \Psi_l(t)\,k^{-n_l}$ for $k\geq k_l(t)$. Nothing is assumed concerning the behavior of $E(k,t)$ at 
intermediate and high wave numbers $k\gg k_l(t)$. In the self similar regime, $k_l(t)$ 
decreases like $\sim\left({\cal A} g t^2\right)^{-1}$ and the mean turbulent kinetic energy 
increases like $\langle {u^{\prime}}^2\rangle\sim\left({\cal A} g t\right)^2$ \cite{rist}. Provided
 that the spectrum  decreases toward high wave numbers up to
 $k=k_{\eta}(t)$ above which it is zero, it can be found that 
\begin{equation}
\langle
 {u^{\prime}}^2\rangle=\int_{k_l}^{k_{\eta}}dk\,
 E(k,t)\sim\frac{\Psi_l(t)}{n_l-1}k_l^{1-n_l}\label{eqinj}
\end{equation} if $n_l>1$ and $k_{\eta}\gg k_l$. We could have
 taken into account an additional and more realistic $\sim k^2$ spectrum
 in the region $k<k_l(t)$ instead of zero but that would only have
 changed the unimportant constant coefficient in front of the above
 result. For the previously depicted self similar
 evolution to occur, using (\ref{eqinj}), the spectrum level at low
 wave number must verify $\Psi_l(t)\sim\left({\cal
 A}g\right)^2\,t^2\,k_l^{n_l-1}\sim \left({\cal
 A}g\right)^{3-n_l}\,t^{4-2n_l}$. The parametrized trajectory of $\lambda$
 in the $k$-$E$ plot can then be determined since $k_{\lambda}(t)=k_l(t)$ and
 $E_{\lambda}(t)=E(k_{\lambda}(t),t)$. After eliminating the variable
 $t$, it is found that the point $\lambda$ must evolve on the curve 
\begin{equation}
E_{\lambda}\sim\left({\cal A}g\right)\,k_{\lambda}^{-2}~,\label{sign}
\end{equation} independently of the slope of the spectrum ($n_l$). This is a universal condition in the
sense that both those who believe in a Kolmogorov scenario $n_l=5/3$ and others will 
agree on. It is objective since it amounts to look for a straight line in the $\log(E)$-$\log(k)$ 
plot whose slope, this time, is unambiguously determined ($-2$). Therefore, such
behavior should be checked whenever an RT flow is said to have reach a
self similar regime. This regime requires large resolution so that the mixing 
zone can expand and reach the self
similar regime before it collides with the simulation box border. Indeed,
large resolution simulations \cite{cook2, cook3, cab1} show
the mark (\ref{sign}) of the self similar regime which starts near the
end. The only
experimental result which shows the evolution of the velocity spectrum
\cite{andr3} also complies with this rule. 

It is now possible to select, among all the simulations cited in the
 introduction, those which have reached the self similar regime. If it is assumed that RT turbulence follows
 Kolmogorov's mechanism \cite{rist,chert1}, the
 turbulent spectrum has the well known form
 $E(k)=C_K\,\epsilon^{2/3}\,k^{-5/3}$. Using Eq.(\ref{eqinj}) with
 $n_l=5/3>1$ and the self similar laws for $k_l(t)$ and $\langle
 {u^{\prime}}^2\rangle$, it is straightforward to conclude
 \cite{chert1} that
 $\epsilon\sim t$ and $k_{\eta}(t)=\left(\epsilon/\nu^3\right)^{1/4}\sim
 t^{1/4}$ and also that $\Psi_l(t)=C_K\,\epsilon(t)^{2/3}\sim
 t^{2/3}$. This conclusion is refuted by recent DNS/LES
 simulations \cite{cook2,cook3} showing the velocity spectrum evolution
 in time. Even though it is not stated in these references, it can
 clearly be noticed that the level of the velocity spectrum at low
 wave number, $\Psi_l(t)$, does not grow as time evolves but
 remains still. This is an important observation which
 constrained the mechanism of RT turbulence.

In order to understand the mechanism, it is worth writing the
 equation governing the evolution of $E(k,t)$ (a generalisation of Lin's equation \cite{sbp})
out of the averaged second moment equation of
 (\ref{equ}) to make the buoyancy production term appear. It can be achieved by multiplying (\ref{equ}) by
 $\bm{u}$. The spectral equation can then be derived by
applying a Fourier transform, $\widetilde{.}$, in the $xy$ plan to the
 resulting equation and
retaining the zero mode contribution. In addition to the rhs of Lin's
equation, we find a spectral buoyancy production term deriving from
$2 {\cal A}g\,\widetilde{u^{\prime}_z c^{\prime}}(0)={\cal A}\,g\,\int
d^2k\,\left[\widetilde{u}_z(k) \widetilde{c}^*(k)+\widetilde{u}^*_z(k)
  \widetilde{c}(k)\right]$ which, assuming homogeneity and isotropy in
the $xy$ plan and phase coherence between $\widetilde{u}_z(k)$ and
$\widetilde{c}(k)$, reduces to $\int dk\,{\cal
  A}g\,E_c^{1/2}(k)E^{1/2}(k)$ since
$\left|\widetilde{u}_z(k)\right|\sim k^{-1/2}E^{1/2}(k)/\sqrt{2\pi}$
and $\left|\widetilde{c}(k)\right|\sim k^{-1/2}E_c^{1/2}(k)/\sqrt{2\pi}$. The resulting generalisation of Lin's spectral equation is therefore 
\begin{equation}
\partial_t E=T(k)-2\nu k^2 E+\beta{\cal A}g E_c^{1/2}E^{1/2}~,\label{eqspec}
\end{equation} where $\beta$ accounts for phase incoherence between
$\widetilde{c}$ and $\widetilde{u}_z$. It
depends on $k$ but remains of order unity. The first term in the rhs ($T$
term) is the so called spectral energy transfer. In the case of a forced
turbulence (Kolmogorov mechanism), this term is negative in the low wave number, to balance
with the production, it is approximately zero throughout the inertial
range and become positive in the dissipative range to balance with
dissipation. 
It accounts for the non linear triad interaction responsible for the
forward cascade in forced
turbulence and must verify $\int_0^{\infty}dk\,T(k)=0$. The second term is the exact contribution of viscosity in the spectral
evolution equation and it is responsible for 
dissipation at high wave number in forced turbulence. The last term, on
the other hand, has never been written
to the author's knowledge. It is this term which makes RT flows and
buoyant flows in general so different. As expected intuitively, it depends on the concentration
spectrum because it is a concentration heterogeneity that induces motion through buoyancy. In order to distinguish the influence
of all three terms, knowledge of the order of magnitude of
$T(k)$ is required. This term has contribution from
$u^{\prime}_j{u^{\prime}_i}^2$ which in spectral language means
$E^{3/2}$ (power counting of $u^{\prime}$ is $3$ which means $3/2$ in
term of $E$) and more precisely
$k^{3/2}E^{3/2}$ for homogeneity reason. The same sort of
argument is used with the pressure term $u^{\prime}_j p^{\prime}$ which
is a non local term  with two sources : pure advection and
buoyancy. Thus, the pressure contribution brings another
$k^{3/2}E^{3/2}$ due to pure advection and a ${\cal A}g
E_c^{1/2}E^{1/2}$ for buoyancy (power counting of
$u^{\prime}$ and $c^{\prime}$ in buoyancy yields $1/2$ for $E$ and $E_c$). And finally,
$\nu\,\partial_j{u^{\prime}_i}^2$ is responsible for a $\nu k^2 E$
  contribution. In a nutshell, $T(k)$ has
  three spectral contributions whose order of magnitude are
  (i) $k^{3/2}E^{3/2}$, (ii) $\nu k^2 E$ and (iii) ${\cal A}\,g
  E_c^{1/2}E^{1/2}$. Thus, the evolution of the velocity spectrum
density in Eq.(\ref{eqspec}) is also controlled by these three
contributions. It is enlightening to draw on a plot (Fig.\ref{zone}) the
predominance domains of the non linearity
(i), viscosity (ii) and buoyancy
(iii) with $k$ on the horizontal axis and $E(k)$ on the vertical axis. 
It is done by equating the three terms two by two.
It gives three boundary lines in a log-log plot which will
be referred to as the non-linearity-buoyancy (NLg), the non-linearity-viscosity 
(NL$\nu$) and the buoyancy-viscosity (g$\nu$) boarders respectively
described by 
\begin{eqnarray}
E_{\mathrm{NLg}}(k,t)&\sim&{\cal A}g\,k^{-3/2}\,E_c(k,t)^{1/2}~,\label{nlg}\\
E_{\mathrm{NL}\nu}(k)&\sim&\nu^2\,k~,\label{nlnu}\\
E_{\mathrm{g}\nu}(k,t)&\sim&\left({\cal A}g\right)^2\nu^{-2}\,k^{-4}\,E_c(k,t)\label{gnu}~.
\end{eqnarray} 
  \begin{figure}
  \begin{center}
  \input{zone.tex}
  \end{center}
  \caption{\label{zone}Predominance diagram of various terms in the
    spectral equation (\ref{eqspec}). In this diagram, the point
    $\lambda$ evolves on the $\lambda$ line following
    $k_l(t)$. The gray curve represents the velocity spectrum
    in the RT self similar regime.}
  \end{figure}
Firstly, $E_c(k,t)$ may not have the same power law at
low wave number (NLg) in (\ref{nlg}) and at high wave number (NL$\nu$) in (\ref{gnu}). In
the self similar regime $\langle{c^{\prime}}^2\rangle$ tends to be a constant \cite{rist}
which will be denoted $c_0^2$. At low wave numbers, by applying Eq.(\ref{eqinj}) to
$\langle{c^{\prime}}^2\rangle$, it is found that $E_c(k,t)$ must vary
like $c_0^2\,k_l^{{n_c}_l-1}(t)\,k^{-{n_c}_l}\sim c_0^2\left({\cal A}g\right)^{1-{n_c}_l}\,t^{2(1-{n_c}_l)}\,k^{-{n_c}_l}$, where ${n_c}_l$ is undetermined yet. 
That is why, in the self similar regime and at
low wave number, Eq. (\ref{nlg}) may be refined as
\begin{equation}
E_{\mathrm{NLg}}(k)\sim c_0\left({\cal
  A}g\right)^{\frac{3-{n_c}_l}{2}}\,t^{1-{n_c}_l}\,k^{-\frac{3+{n_c}_l}{2}}~.\label{nlg2}
\end{equation} It is to
be noticed that when ${n_c}_l=1$, the previous time evolution is
changed to $\log(t)$. Secondly, it is worth noticing that $E_{\mathrm{NL}\nu}(k)\sim\nu^2\,k$ 
is independant of the concentration spectrum $E_c(k,t)$, that it does not depend on time and that it is the exact 
line where the inertial cascade vanishes and where
  dissipation starts acting in the Kolmogorov mechanism. 

It is now possible to describe the evolution of the
  velocity spectrum in an RT mixing layer. Let us imagine that a peaked initial condition is chosen at a wave
  number below $\left({\cal A}g/\nu^2\right)^{1/3}$. Spontaneous RT
  can then occur. The closest to $\left({\cal
  A}g/\nu^2\right)^{1/3}$, the fastest is the linear growth. The spectrum
  grows until it reaches the NL domain where mode
coupling can start. At the beginning, since both $E(k)$ and $E_c(k)$
are peaked around the initial wave number, so is the buoyancy
production which then acts like a narrow band forcing. A Kolmogorov
spectrum can then settle between the initial wave number and
NL$\nu$ (where energy is dissipated). This is precisely what is
observed after the linear growth regime in the mixing phase. As time
goes by, both spectrums spread over a wider range of wave numbers and when
mixing is established, production becomes broad band and spectral energy transfer must
balance with buoyancy at the lowest wave number. That is why, the velocity spectrum
 lies alongside $E_{\mathrm{NLg}}(k)$ (see Fig.\ref{zone}), given by (\ref{nlg}) and also
 (\ref{nlg2}) in the self similar RT case. 

This balance
 mechanism is observed for Rayleigh-B\'enard (RB) flows for it
  explains its velocity and concentration spectrum layout. RT and RB flows 
are both governed by the
 same set of equation (concentration is replaced by temperature in RB)
  but boundary conditions in RB are independent of time and so are the
 velocity and concentration spectrum. At low wave numbers,
 $\partial_t{c^{\prime}}^2\sim
 -\partial_j\left(u^{\prime}_j{c^{\prime}}^2\right)$  which in
 spectral language approximately means $\partial_t
 E_c\sim\partial_k\left(k^{5/2}E^{1/2}E_c\right)$. If $E_c$ is
 independent of time at low wave numbers in the self similar regime,
 that means the spectral transfer of concentration $\sim
 k^{5/2}E^{1/2}E_c$ must be independent of $k$, that is to say
 $\frac{5}{2}-\frac{n_l}{2}-{n_c}_l=0$ for RB flows (power counting of
 $k$). Moreover, since (\ref{nlg}) must be
 valid in RB flows (Eq. (\ref{nlg2}) is not valid in this case
 because RT boundary conditions have been used to derive it) the relation
 $-n_l=-\frac{3}{2}-\frac{{n_c}_l}{2}$  must also be true. Together
 with the previous relation it is then found that
 $n_l=\frac{11}{5}=2.2$ and ${n_c}_l=\frac{7}{5}=1.4$, which is exactly
 the Bolgiano-Obukov (BO) \cite{bolg} scaling found in RB flows \cite{rb, nature}.
 This mechanism is also corroborated by  experiment \cite{andr3} where
  velocity spectrum level at low wave numbers can be seen to decrease
 in time although the turbulent kinetic energy increases. That means,
  (\ref{nlg2}), that ${n_c}_l\gtrsim 1$, which is confirmed by all
  numerical and experimental results, and that, once again,
  Kolmogorov mechanism is ruled out since it predicts a velocity
  spectrum level increase which would require ${n_c}_l=1/3<1$. 

In the high wave numbers, depending on the value of the Schmidt number, two
  things can happen. $Sc\lesssim 1$ means that buoyancy production
  cannot exist below $\nu^2\,k$ and cannot balance with viscosity to
  create $g\nu$ (see Fig.\ref{zone}). In this case, the velocity spectrum exponentially decreases
  after it goes through $NL\nu$. Otherwise, if $Sc\gg 1$ it is possible to prove
  that an equilibrium between buoyancy and viscosity exists by performing a
 linear analysis of (\ref{eqc},\ref{equ},\ref{eqdiv}) with
  $c(\bm{x},t)=c_0+c_1(\bm{k},t)\,e^{-i\bm{k}\bm{x}}$ and $\bm{u}(\bm{x},t)=\bm{u}_0+\bm{u}_1(\bm{k},t)\,e^{-i\bm{k}\bm{x}}$. This solution can
be thought of as the low wave number contributions of $\bm{u}$ and
$c$ with perturbation corresponding to higher wave number
  turbulent fluctuations. They can be plugged in equations
  (\ref{eqc},\ref{equ},\ref{eqdiv}) and after some algebra it is found
  that
\begin{eqnarray}
c_1(\bm{k},t)&=&c_1^0(\bm{k})\,e^{(i\bm{u}_0\bm{k}-\kappa k^2)t},\\
{u_1}_i(\bm{k},t)&\approx&\frac{g_j}{\nu\,k^2}\left(\delta_{ij}-\frac{k_i
  k_j}{k^2}\right)
  c_1^0(\bm{k})\,e^{(i\bm{u}_0\bm{k}-\kappa
      k^2)t},\label{lin2}
\end{eqnarray} in the high Schmidt number limit. As a result and by definition of
$E(k)$ and $E_c(k)$ it is then found
\begin{equation}
E(k,t)=\frac{8 \left({\cal A}\,g\right)^2}{\nu^2 k^4}E_c(k,t)~,
\end{equation}in agreement with Eq.(\ref{gnu}). Therefore, the velocity
spectrum coincide with $E_{g\nu}(k)$ in the high Schmidt number limit
(see Fig.\ref{zone}). 

As a conclusion, it can be stated that self
similarity hypothesis together with equilibrium of spectral energy
transfer with buoyancy at low wave numbers constrained velocity
spectrum in a way incompatible with Kolmogorov mechanism. It does not
provide the exact slope of the concentration spectrum ${n_c}_l$ at low
wave numbers but assuming that it lies between $1$ and $2$ we obtain
$2\leq n_l\leq 2.5$. Recent high resolution simulations, show that the
velocity spectrum level remains close to the $\lambda$ line which
would be in favor of ${n_c}_l\approx 1$ and $n_l\approx 2$ as a
result. The most important result, here, is that velocity spectrum at
low wave numbers is shown to be very sensitive to concentration
spectrum. Nevertheless, a Kolmogorov mechanism is not ruled out at
intermediate wave numbers. It can be a transition process from low wave
numbers to high wave numbers as in the BO mechanism. Most of the gravitational energy is injected through buoyancy at low wave numbers up to a critical wave number. Above it, injection of energy is negligible and a Kolmogorov mechanism is possible. In this case,
it does not involve that the turbulent dissipation $\epsilon=2\nu\int
dk\,k^2 E(k)$ should vary like $\epsilon\sim t$. As mentioned in
\cite{rist}, this common belief that ``the cascade rate is
[...] equal to dissipation'' in the self similar regime is always
presupposed although it is a strong hypothesis for buoyant flows. It
does not have to hold considering the importance of the buoyant
production. It would not be in contracdiction with turbulent kinetic
energy conservation
$\partial_t\langle{u^{\prime}_i}^2\rangle=\Pi-\epsilon$. Indeed, since buoyancy
production $\Pi\sim t$ and
$\partial_t\langle{u^{\prime}_i}^2\rangle\sim t$, dissipation could
vary slowlier without threatening energy conservation in the self similar regime.

\end{document}